\documentclass[10pt,letterpaper]{article}
\usepackage{opex3}
\usepackage{enumitem}
\usepackage{amssymb}
\usepackage{cite}

\DeclareFontFamily{U}{euc}{}
\DeclareFontShape{U}{euc}{m}{n}{<-6>eurm5<6-8>eurm7<8->eurm10}{}%
\DeclareSymbolFont{AMSc}{U}{euc}{m}{n}
\DeclareMathSymbol{\umu}{\mathord}{AMSc}{"16}

\begin{document}

\title{Arm-length stabilisation for interferometric gravitational-wave
  detectors using frequency-doubled auxiliary lasers}

\author{Adam J.\ Mullavey,$^{1}$ Bram J.\ J.\ Slagmolen,$^1$ John
  Miller,$^{1,*}$\\ Matthew Evans,$^2$ Peter Fritschel,$^2$ Daniel
  Sigg,$^3$ Sam J.~Waldman,$^2$\\ Daniel A.\ Shaddock,$^1$ and
  David~E.~M$^\mathrm{c}$Clelland$^{1}$}

\address{ $^1$Centre for Gravitational Physics, The Australian
  National University,\\
  Canberra, ACT, 0200, AUSTRALIA\\
  $^2$LIGO Laboratory, Massachusetts Institute of Technology,\\
  185 Albany St, Cambridge, MA 02139, USA\\
  $^3$LIGO Hanford Observatory, PO Box 159,
  Richland, WA 99352, USA}

\email{*john.miller@anu.edu.au}

\begin{abstract}
  Residual motion of the arm cavity mirrors is expected to prove one
  of the principal impediments to systematic lock acquisition in
  advanced gravitational-wave interferometers. We present a technique
  which overcomes this problem by employing auxiliary lasers at twice
  the fundamental measurement frequency to pre-stabilise the arm
  cavities' lengths. Applying this approach, we reduce the apparent
  length noise of a 1.3~m long, independently suspended Fabry-Perot
  cavity to 30~pm~rms and successfully transfer longitudinal
  control of the system from the auxiliary laser to the measurement
  laser.
\end{abstract}





\section{Introduction}
Direct detection of gravitational radiation, predicted by Einstein's
general theory of relativity, remains one of the most exciting
challenges in experimental physics. Due to their relatively weak
interaction with matter, gravitational waves promise to allow
exploration of hitherto inaccessible processes and epochs
\cite{Cutler2002}. Unfortunately, this weak coupling also hinders
detection with strain amplitudes at the Earth estimated to be
${\scriptstyle \lesssim}10^{-21}$. Nevertheless, the network of
advanced gravitational wave detectors currently under construction
\cite{Luck2010,Kuroda2010,Harry2010,Accadia2011} is widely expected to
operate with sufficient sensitivity to observe several events per year
(see e.g.~\cite{Abadie2010D}).

\begin{figure}[htbp!]
\centerline{\includegraphics[width=7.5cm]{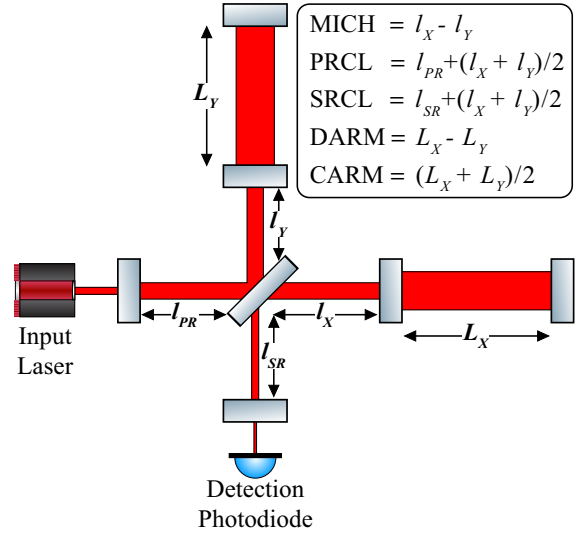}}
\caption{(Colour online) Schematic of a contemporary
  gravitational-wave interferometer indicating primary length degrees
  of freedom. In this work MICH, PRCL and SRCL are described as
  \emph{central} degrees of freedom.  The arms have lengths $L_{X,Y}$
  of order 1~km; the other cavities, PRCL and SRCL, are significantly
  shorter (${\scriptstyle \lesssim}$50~m).}
\label{fig:IFO}
\end{figure}
Modern gravitational-wave detectors are Michelson-style
interferometers, enhanced by the addition of resonant cavities at
their inputs, outputs and, generally, in each of their arms (see
Fig.~\ref{fig:IFO}). When all of these cavities are held within their
respective linewidths by interferometer control systems we say that
the interferometer is \emph{locked}. When the interferometer is not
locked no meaningful scientific data can be recorded. Due to
interactions between the optical cavities, lock acquisition is a
non-trivial problem.

The second generation of interferometric gravitational-wave detectors
will employ higher finesse (narrower linewidth) arm cavities. Recent
investigations indicate that it is these arm cavities which will pose
the greatest challenges during the lock acquisition process
\cite{Miyakawa06,WardThesis}. In this work we develop a tool, an \emph{arm-length
  stabilisation system} or \emph{ALS}, to address these challenges.

\section{Arm-length stabilisation}
The length degrees of freedom of all gravitational-wave
interferometers are controlled using an extension of the
Pound-Drever-Hall (PDH) technique \cite{Drever83} -- radio-frequency
phase-modulation sidebands are impressed upon the input laser light at
multiple frequencies and the circulating field is detected and
demodulated at selected interferometer output ports \cite{Miyakawa06,WardThesis}.

The resonant state of the modulation sidebands, the demodulation
frequencies and phases, and the macroscopic cavity lengths are all
carefully chosen to provide low-noise sensing signals for each of the
degrees of freedom when the interferometer is locked. In particular,
the modulation frequencies are chosen such that the control sidebands
do not resonate inside the arm cavities.

Due to the optical couplings between the various cavities, these
detection schemes do not always provide reliable sensing signals
during lock acquisition. In this respect, the arm cavities are
singularly troublesome.

Advanced gravitational-wave interferometers utilise multi-stage
seismic isolation systems which offer excellent performance above
$\sim$1~Hz (see e.g. \cite{Robertson04bb}); however, it remains
difficult to suppress noise from lower frequency sources (e.g.\
double-frequency microseism). Consequently, the residual arm cavity
length noise is expected to be $\sim$1 $\umu$m rms, 1000 times greater
than a typical arm cavity's linewidth ($\sim$1~nm). Displacements of
this magnitude are problematic for two reasons:

Firstly, the carrier and control sideband fields will occasionally
become resonant in the arms. Sensing signals for the central degrees
of freedom are derived from the interaction between the carrier and
sidebands or between the sidebands themselves. When one component of
either pair becomes resonant, control signals for the central degrees
of freedom become invalid.

\newpage
Secondly, for reasons of noise, second generation detectors will
employ significantly weaker test mass actuators (see
e.g.~\cite{Miller2011A}) and utilise heavier test masses
($\sim$40~kg). As a result of these choices, the test mass actuators
will often lack sufficient authority to gain control over the arm
cavities when they are freely swinging.

Although it is possible to acquire lock under these conditions, this
acquisition cannot be realised in a repeatable, systematic manner. The
goals of the arm-length stabilisation system are therefore twofold:
\begin{enumerate}
\renewcommand{\labelenumi}{\alph{enumi})}
 \setlength{\itemsep}{5pt}%
 \setlength{\parskip}{0pt}%
\item Maintain both arm cavities at a fixed offset from resonance so
  that the central degrees of freedom may be locked without
  obstruction.
\item Reduce rms cavity motion to within one linewidth ($\sim$1~nm) so
  that arm cavity lock acquisition signals can be usefully applied.
\end{enumerate}
Implicit in these requirements is the ability to methodically remove
the fixed offset from arm resonance to arrive at a state where the arm
cavity acquisition signals are valid.

\section{Technique}
We now describe the approach adopted to achieve the above goals,
providing a general description of the strategy applied followed by
explanatory details concerning one possible practical
implementation. Compared to other techniques considered for arm-length
stabilisation, this approach relies on proven technologies, offers
greater sensitivity \cite{Shaddock07} and is less invasive
\cite{Drever2002}.

\subsection{Design philosophy -- dual-wavelength locking}
An additional \emph{auxiliary} laser is placed behind each end test
mass. These lasers are independently locked to their respective arm
cavities by actuating on the lasers' frequencies, circumventing the
weak test mass actuators.

By comparing the frequencies of the auxiliary lasers to the frequency
of the main interferometer's pre-stabilised laser (PSL), one can
construct ALS signals describing the offset of the PSL from resonance
in the arms. Outside of the cavity linewidth, conventional length
sensing signals are often non-linear and cannot be used to effect
closed-loop control.  In contrast, these ALS signals remain valid even
when the PSL is far removed from resonance; hence they can be used to
actively stabilise and adjust the detunings of the arms during lock
acquisition by actuating on the end test masses.

To avoid cross-coupling between main interferometer and arm-length
stabilisation signals, the auxiliary lasers operate at 532~nm. This
wavelength was chosen for its harmonic relationship to the wavelength
of the PSL (1064~nm). The use of two distinct wavelengths demands that
the arm cavity mirror coatings be dichroic. The choice of cavity
finesse (i.e.\ mirror reflectivities) at 532~nm is relatively
free. Low values ease auxiliary laser lock acquisition whilst higher
values provide improved mode filtering and noise performance. A
finesse of around 100 represents a reasonable compromise. Dichroic
mirrors compatible with this specification are not expected to
increase the observed mirror thermal noise significantly
\cite{Principe2010, Villar2010}.

\subsection{Practical implementation}
\label{sec:implementation}
\begin{figure}[htbp!]
\centering\includegraphics[width=9cm]{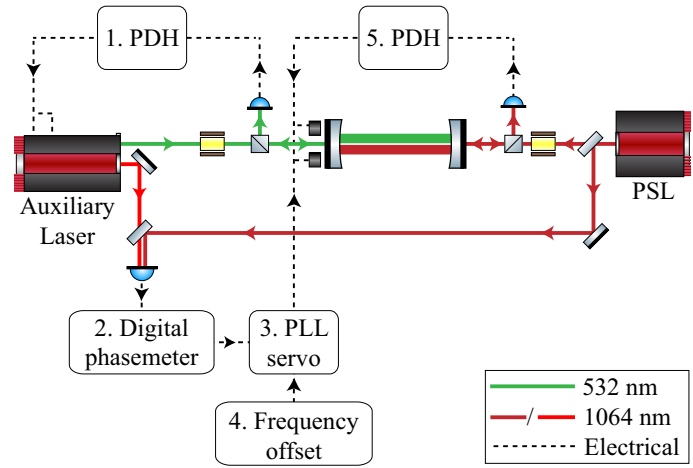}
\caption{(Colour online) Schematic of the arm-length stabilisation
  system. The numbering indicates the flow of the lock acquisition
  process and corresponds to the enumerated list below.}
\label{fig:setup}
\end{figure}
We now proceed through our realisation of this arm-length
stabilisation strategy sequentially (see Fig.~\ref{fig:setup}). For
clarity, we consider only a single resonant cavity, representing one
arm of an advanced interferometer.

\begin{enumerate}[topsep=0mm, labelindent=0em,  leftmargin=*]
\item The 532~nm output of a dual-wavelength (1064~nm and 532~nm)
  auxiliary laser is locked to the arm cavity using the PDH
  technique. This wide-bandwidth ($>$10~kHz) control loop provides the
  reference measurement of the arm's resonant frequency. The auxiliary
  laser remains tightly locked to the arm cavity at all times whilst
  the ALS system is active.

\item The frequency of the auxiliary laser is subsequently compared to
  that of the PSL by measuring the frequency of their heterodyne beat
  note using a LISA-like digital phasemeter \cite{Shaddock2006}. This
  comparison is made at 1064~nm using the auxiliary laser's second
  output (which has a constant phase relationship with the 532~nm
  beam). The beat-note frequency indicates how far the PSL beam is
  from resonating in the arm cavity. The extensive linear range of
  this heterodyne measurement, compared to conventional PDH-based
  sensors, is the key feature of the ALS system.

  In an operational gravitational-wave detector this measurement
  necessitates the transfer of a frequency reference through
  $\sim$4~km of optical fibre (from the PSL to the auxiliary laser or
  vice versa). Well-established techniques to cancel noise induced by
  fibre transmission exist (e.g.~\cite{Ma94,Mullavey2010}).

\item The output from the digital phasemeter is then used to offset
  phase-lock the auxiliary laser to the PSL. Feedback signals are
  applied to the arm cavity's end test mass, effectively suppressing
  the cavity's motion relative to the PSL and stabilising its offset
  from resonance.

\item By adjusting the offset frequency of the phase-locked loop
  (PLL), the detuning of the PSL from resonance can be actively
  controlled (see Fig.~\ref{fig:sweep}).

  In a long-baseline interferometer, this feature can be employed to
  hold the arm cavities at a fixed offset from resonance, allowing the
  central degrees of freedom to be locked without disturbance.

\item The offset of the arm from resonance may now be reduced in a
  methodical fashion, bringing the cavity into a region where
  interferometer acquisition signals can be activated. It is also
  feasible that the ALS system could bring the arms fully onto
  resonance so that low-noise, operating-mode control signals can be
  engaged directly. This second approach was simulated in our
  experiment whereby control over the arm cavity's length was
  transferred from the ALS system directly to a PSL PDH error signal
  captured in reflection (see Fig.~\ref{fig:handoff}).

  \newpage
  After control is transferred to the main interferometer,
  the ALS system may be stood down so that it does not introduce any
  additional noise to the instrument. Alternatively, the ALS system
  may be retained as a powerful diagnostic tool. For example, we
  anticipate that the phase-locked auxiliary lasers will be able to
  provide accurate measurements of arm cavity alignment, g-factor and
  absolute length. Whether these measurements can be made on-line
  remains to be determined.

\end{enumerate}
Generalisation of this scheme to a two-arm interferometer involves
combining the ALS signals from each arm (either optically or
electronically) to construct signals which align with the notional
common-arm and differential-arm degrees of freedom used in
interferometer control (CARM and DARM in Fig.~\ref{fig:IFO}).

For further details on the arm-length stabilisation concept and its
role in the lock acquisition process see \cite{Slagmolen2011}.

\section{Experimental test}
In order to validate the fundamental approach discussed above, a
laboratory-scale proof-of-principle experiment was carried out at The
Australian National University's Centre for Gravitational Physics. To
concentrate on the novel aspects of the arm-length stabilisation
system, we again examined only a single optical resonator.

Our 1.3~m long cavity was formed from two single-stage piano-wire
suspension systems known as `Tip-Tilts' \cite{Slagmolen10C}; it had a
g-factor of 0.46 and measured finesses of 300 at 1064~nm and 100 at
532~nm. (For comparison, the Advanced LIGO detectors are expected to
have finesses of approximately 450 and 100.) The dichroic cavity
mirrors were controlled by coil-magnet actuators via an Advanced LIGO
digital control system. The role of the PSL was played by a standard
diode-pumped solid-state laser (JDSU NPRO 126); the auxiliary laser
was an Innolight Prometheus.

\begin{figure}[htbp!]
\centering\includegraphics[width=9cm]{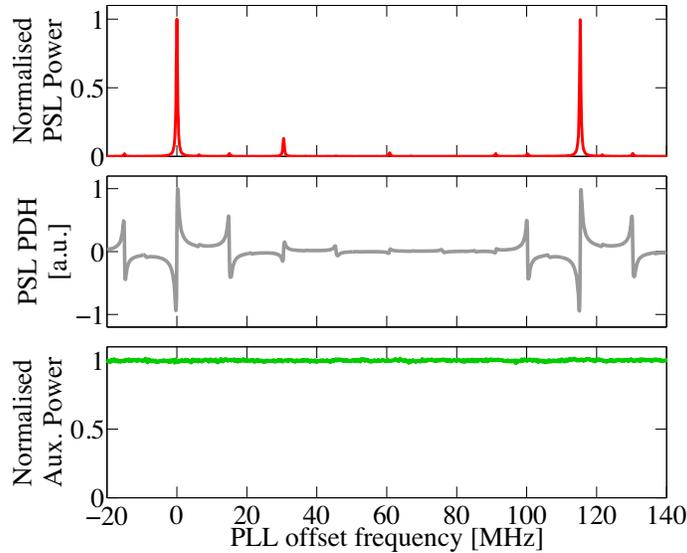}
\caption{(Colour online) Systematic cavity detuning over more than one
  free spectral range using the arm-length stabilisation system. Top
  -- Normalised cavity transmission at the wavelength of the
  measurement laser (1064~nm); Middle -- Pound-Drever-Hall signal
  generated from the measurement laser alone; Bottom -- Normalised
  cavity transmission at the wavelength of the auxiliary laser
  (532~nm).}
\label{fig:sweep}
\end{figure}
In Fig.~\ref{fig:sweep} we conclusively demonstrate the technique's
capacity to explore the full range of arm cavity detunings. With the
ALS system active, the offset frequency of the phase-locked loop (item
4 in Fig.~\ref{fig:setup}) was swept linearly over more than one free
spectral range. Since the auxiliary laser is securely locked to the
arm cavity, this offset frequency directly controls the detuning of
the PSL from resonance.

In a gravitational-wave interferometer this capability would permit us
to maintain a specified detuning, away from any undesirable
resonances, allowing the central degrees of freedom to be easily
locked, thus meeting the first ALS goal. The extent to which the
specified detuning is `fixed' will be explored below.

\begin{figure}[htbp!]
\centering\includegraphics[width=9cm]{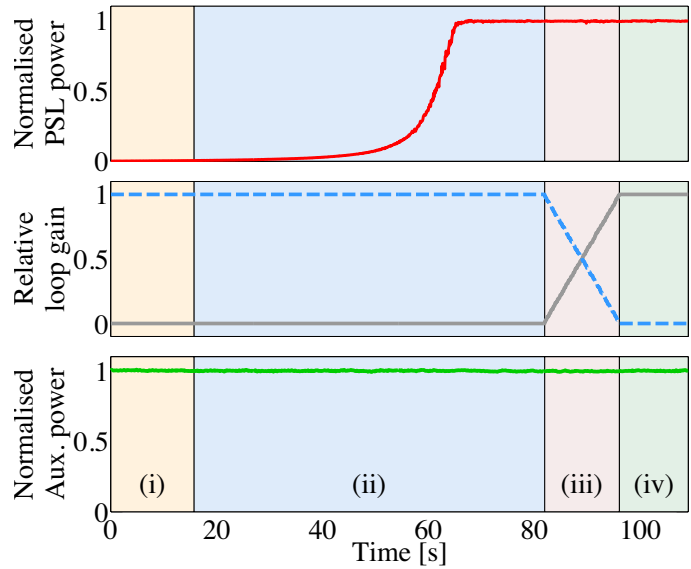}
\caption{(Colour online) Transfer of arm cavity length control from
  the arm-length stabilisation system to signals derived solely from
  the measurement laser. Top -- Normalised cavity transmission at the
  wavelength of the measurement laser (1064~nm); Middle -- Relative
  gain of arm-length stabilisation (blue dashed) and measurement laser
  (grey) control signals; Bottom -- Normalised cavity transmission at
  the wavelength of the auxiliary laser (532~nm). The division of the
  axes into four regions is discussed in the main text. The timescale
  of this handover does not represent the limit of system
  performance.}
\label{fig:handoff}
\end{figure}

Complete command over arm cavity detuning also allows us to satisfy
the implicit goal of manoeuvring the cavity system from a stable
off-resonance state to a position where acquisition signals become
meaningful. A typical handover from ALS to PSL control signals is
shown in Fig.~\ref{fig:handoff}.

The axes are divided into four shaded regions, representing different
stages of the transfer:
\begin{enumerate}
\renewcommand{\labelenumi}{(\roman{enumi})}
 \setlength{\itemsep}{3pt}%
 \setlength{\parskip}{0pt}%
\item The cavity is initially stabilised at a point far from
  resonance. The detuning is reduced in an orderly fashion by
  adjusting the offset frequency of the phase-locked loop.

\item The cavity approaches resonance, circulating power begins to
  increase and PSL-based control signals become viable.

\item Control over the cavity length is transferred to the PSL. As
  both PSL and ALS systems actuate on the cavity's end test mass, this
  handover is realised simply by tuning the relative gain of the two
  feedback loops.

\item The cavity is under the control of PSL signals alone.
\end{enumerate}

\begin{figure}[htbp!]
\centering\includegraphics[width=9cm]{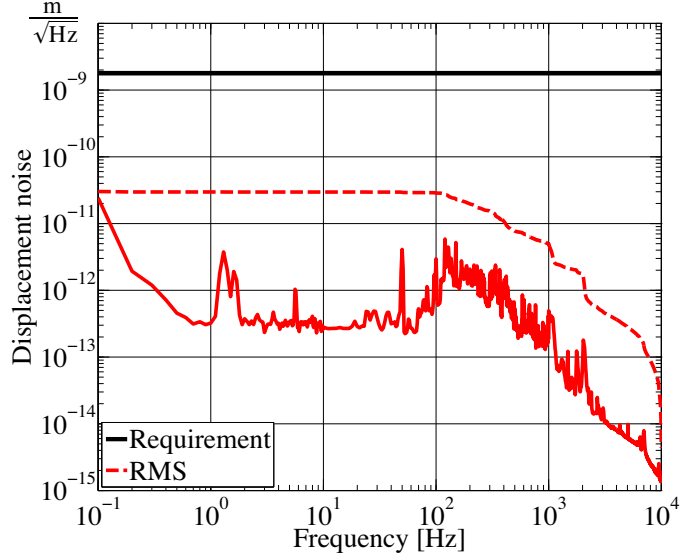}
\caption{(Colour online) Residual cavity displacement noise relative
  to the measurement laser with arm-length stabilisation system
  active. The integrated rms noise (dashed line) is within one
  full-width-half-maximum cavity linewidth (solid horizontal line) as
  required. Data shown in this figure were taken with our optical
  table's pneumatic vibration isolators activated. The prominent
  features around 1~Hz are due to the mechanical resonances of this
  system. All other presented data were recorded with this isolation
  system turned off.}
\label{fig:results}
\end{figure}
Recall that the second goal of the arm-length stabilisation system is
to reduce the rms displacement noise of the arm cavity, relative to
the PSL, to within one linewidth.  The full-width-half-maximum-power
cavity linewidth is given by
\begin{equation}
  \label{eq:linewidth}
 \Delta_\mathrm{FWHM}=\left\{
   \begin{array}{rl}
      \lambda/(2\mathcal{F})&\mbox{[m]}\\
      c/(2L\mathcal{F})&\mbox{[Hz]}
    \end{array} \right.,
\end{equation}
where $\lambda$ is the laser wavelength, $\mathcal{F}$ is the cavity
finesse, $c$ is the speed of light and $L$ is the cavity length. For
our parameters the cavity linewidth is approximately 1.8~nm,
comparable to the Advanced LIGO value of 1.2~nm.

This specification was tested by tuning the cavity onto resonance
using the arm-length stabilisation system and employing the PSL PDH
measurement as an out-of-loop sensor. The resulting amplitude spectral
density is shown in Fig.~\ref{fig:results}. The integrated rms motion
(dashed line) was found to be 30.2~pm, comfortably meeting the
cavity-linewidth requirement (solid horizontal line).

Figure~\ref{fig:results} also describes the stability of \emph{any}
offset from resonance (e.g.\ that introduced when locking the central
degrees of freedom) as the performance of the ALS system does not vary
as a function of arm cavity detuning.

Combined, the above findings demonstrate the validity of arm-length
stabilisation approaches based on frequency-doubled auxiliary
lasers. This positive result should, nevertheless, be considered in
context. Any extrapolation of the work presented here to a
kilometre-scale interferometer will require the differences in
environment, test mass actuation and optical configuration to be
addressed. However, recent simulation work predicts that, taking these
differences into account, the linewidth specification can still be met
\cite{Slagmolen2011}.

\section{Discussion}
The results presented in Fig.~\ref{fig:results} reveal an increase in
noise at low frequencies ($<$1~Hz). It is suspected that this roll-up
is due to a combination of spurious amplitude modulation introduced by
our electro-optic modulators and mirror alignment fluctuations (our
cavity was not instrumented with any auto-alignment systems). Both
effects can be mitigated should it be found necessary; however, the
measured noise is a factor of 60 below the linewidth requirement and
smooth cavity tuning and efficient control transfer were achievable at
all times.

For ideal operation, the arm-length stabilisation strategy described
herein demands that both the PSL and the auxiliary laser sense
identical cavity lengths. In practice, a number of
wavelength-dependent effects limit the extent to which this is
possible. For example, the two beams have different spot sizes on the
mirrors, field penetrations into the dichroic coatings and
susceptibilities to cavity misalignment. In our (bench-top) experiment
air turbulence was identified as a significant noise source. This
effect may have been exacerbated by the different mode volumes
occupied by the two lasers within the cavity.

\section{ Conclusions}
In this investigation we have developed the general method of
arm-length stabilisation based on auxiliary lasers. We have
demonstrated the viability of this approach using a single cavity,
stabilising its residual motion to within one cavity linewidth.

Our method is described as a series of key measurements. Each of these
measurements can be made using several proven techniques, allowing the
scheme to be easily modified without reducing capability.

A conceptually identical arm-length stabilisation system, based on
frequency-doubled auxiliary lasers, has now been selected as a
baseline technology for Advanced LIGO. Testing of this scheme on a
fully-suspended, dual-recycled interferometer is underway at the
California Institute of Technology.

The integration of the ideas introduced here into the Advanced LIGO
length sensing and control architecture will not be without
challenges. However, an effective arm-length stabilisation system
would, for the first time, decouple the arm cavities from the central
degrees of freedom and enable global control to be achieved from the
start of a repeatable and diagnosable lock acquisition sequence.

\section*{Acknowledgements}
The authors gratefully acknowledge the assistance of the LIGO
Interferometer Sensing and Control working group. They also thank
Kenneth A.~Strain for useful suggestions during the preparation of
this manuscript and Timothy T.-Y.\ Lam for valuable experimental
assistance. Aspects of Fig.~\ref{fig:setup} were created using
Component Library v.3 by A.~Franzen. This work was supported by the
Australian Research Council. JM is the recipient of an Australian
Research Council Post Doctoral Fellowship (DP110103472). LIGO was
constructed by the California Institute of Technology and
Massachusetts Institute of Technology with funding from the National
Science Foundation and operates under cooperative agreement
PHY-0757058. This paper has been assigned LIGO Laboratory document
number P1100134.
\end{document}